\documentclass[aps,showpacs,amsfonts,amsfonts]{revtex4}
\usepackage{feynmf}

\newcommand{\setval}{\fmfset{wiggly_len}{1.5mm}\fmfset{arrow_len}{1.5mm}
\fmfset{arrow_ang}{13}\fmfset{dash_len}{1.5mm}\fmfpen{0.125mm}
\fmfset{dot_size}{0.8thick}}

\def\cm#1{}

\newcommand{\nc}{\newcommand}
\nc{\p}{\partial}
\nc{\f}[2]{\frac{#1}{#2}}
\nc{\beq}{\begin{eqnarray}}
\nc{\eeq}{\end{eqnarray}}
\nc{\bea}{\begin{eqnarray}}
\nc{\eea}{\end{eqnarray}}
\nc{\nn}{\nonumber}
\nc{\Zm}{Z_{m^2}}
\nc{\Zg}{Z_{g}}
\nc{\Zphi}{Z_{\phi}}
\nc{\Ze}{Z_{e}}
\nc{\Za}{Z_{A}}
\nc{\Zv}{Z_{v}}
\nc{\ep}{\varepsilon}
\nc{\mubar}{\bar{\mu}}
\nc{\lna}{\ln\left(\f{m^2}{\mubar^2}\right)}
\nc{\lnb}{\ln^2\left(\f{m^2}{\mubar^2}\right)}
\nc{\lnc}{\ln^3\left(\f{m^2}{\mubar^2}\right)}
\nc{\ga}{\gamma}
\nc{\gv}{\ga_v}
\nc{\ze}{\zeta}

\begin{document}

\bibliographystyle{apsrev}

\title{Three-Loop Ground-State Energy of O($N$)-Symmetric Ginzburg-Landau
Theory Above $T_c$ in $4-\ep$ Dimensions with Minimal Subtraction}

\author{Boris Kastening and Hagen Kleinert}
\affiliation{Institut f\"ur Theoretische Physik, Freie Universit\"at Berlin,
Arnimallee 14, D-14195 Berlin, Germany}
\author{Bruno Van den Bossche}
\affiliation{Physique Nucl\'eaire Th\'eorique, B5,
Universit\'e de Li\`ege Sart-Tilman, 4000 Li\`ege, Belgium\\
and Institut f\"ur Theoretische Physik, Freie Universit\"at Berlin,
Arnimallee 14, D-14195 Berlin, Germany}

\begin{abstract}
As a step towards deriving universal amplitude ratios of the
superconductive phase transition we calculate the vacuum energy density
in the symmetric phase of O($N$)-symmetric scalar QED in $D=4-\ep$
dimensions in an $\ep$ expansion using the minimal subtraction scheme
commonly denoted by $\overline{\mbox{MS}}$.
From the diverging parts of the diagrams, we obtain the renormalization
constant of the vacuum $\Zv$ which also contains information on the
critical exponent $\alpha$ of the specific heat.
As a side result, we use an earlier two-loop calculation of the effective
potential \cite{twoloops} to determine the renormalization constant of
the scalar field $\Zphi$ up to two loops.
\end{abstract}

\maketitle

\section{Introduction}

One of the most intriguing problems in the physics of critical phenomena
is a theoretical understanding of the superconductive phase transition
within the renormalization group approach.
A first discussion was given in 1974 by Halperin, Lubensky, and Ma
\cite{HLM1} on the basis of the Ginzburg-Landau or U$(1)$ Abelian-Higgs
model, in $4-\epsilon$ dimensions, generalizing a similar four-dimensional
analysis of Coleman and Weinberg \cite{CW}.
In a one-loop approximation, they did not find an infrared-stable fixed
point, and in spite of much effort it is still unclear whether a
higher-loop renormalization group analysis would be capable  of explaining
the existence of a critical point in $4-\ep$ dimensions.
Experimentally, this existence has been confirmed only recently with the
advent of high-$T_c$ superconductors.
In conventional superconductors, the Ginzburg criterion \cite{GB}, or the
more relevant criterion for the size of phase fluctuations \cite{KK},
predicted  a too  small temperature interval for the critical regime
to observe anything beyond mean-field behavior.
Evidence had so far come only from Monte-Carlo simulations \cite{MCS}
and an analogy with smectic-nematic transitions in liquid crystal
\cite{NS}.
Only by artificially allowing for an unphysically large number of replica
$n$ of the complex field $\phi$, larger than $365$, has it been possible
to stabilize the renormalization flow in $4-\ep$ dimensions.
Historically, a dual disorder formulation of the Ginzburg-Landau model
brought in 1982 the first theoretical proof for the existence of a
tricritical point at a Ginzburg parameter $\sqrt{2}\kappa\approx0.77$,
a material parameter characterizing the ratio between magnetic and coherence
length scales \cite{Kleinert,KleinertBook}.
This prediction was recently confirmed by extensive Monte-Carlo simulations
\cite{Sudbo}.

The confusing situation in the Ginzburg-Landau model certainly requires
further investigation in higher loop approximations.
So far, two loop renormalization group calculations in $4-\ep$
dimensions have not yet produced  satisfactory results
\cite{Tess,Folk}.
Analyses in $d=3$ dimensions \`a la Parisi have also left many open
questions \cite{deCalan1}.
An interesting observation was made by Nogueira \cite{Nogueira2}, that
an anomalous momentum instability below $T_c$ may be responsible for
the unusual resistance of the superconductive transition to theory.
Hope for a better understanding has also been raised by a recent
renormalization group study in $d=3$ dimensions performed for the first
time {\em below\/} $T_c$ where a fixed point has been found at the one-loop
level \cite{KlNo}.

Once an infrared-stable  fixed point is located, it will determine
critical exponents and amplitude ratios.
The former can be extracted from the perturbative expansions of
the renormalization constants corresponding to coupling, mass and
wave function renormalization.
The latter also requires the calculation of finite parts of certain
quantities.
For instance,\ the amplitude ratios for the specific heat can be obtained
by computing both the divergent and finite parts of the vacuum energy
in the symmetric and in the symmetry-broken cases.

Our work will provide analytic results for the divergent as well as
finite parts of the vacuum energy up to three loops in the symmetric case.
While the divergent part is the same in the symmetry-broken case, the
determination of the finite part in that case is left for future work.

We work with the so-called modified minimal subtraction scheme,
denoted by $\overline{\mbox{MS}}$ \cite{KSF}.
The singularities are collected in the dimensionless renormalization
constant  $Z_v$.
In the course of the calculations, we also recover information about the
other renormalization constants of the theory.
This may be viewed as a cross check of our work.

Since the theory has so far only a fixed point for large $n>365$,
we shall keep an arbitrary number of replica in the theory,
the physical case being $n=1$.
The $n$ complex fields are coupled minimally to an Abelian gauge field
which describes magnetism.
The $N=2n$ real and imaginary parts of the fields are assumed to have
an O($N$)-symmetric quartic self-interaction.

\section{Model}
\label{model}
The Lagrangian density to be studied contains $n=N/2$ complex scalar
fields $\phi_B$ coupled to the magnetic vector potential $A_{B\mu}$ and
reads, with a covariant gauge fixing,
\beq
\label{EQqedlag}
{\cal L}=|D_{B\mu}\phi_B|^2+m^2_B|\phi_B|^2+\f{g_B}{4}|\phi_B|^4
+\f{1}{4}F_{B\mu \nu}^2+\f{1}{2\alpha}(\p_{\mu} A_{B\mu})^2,
\eeq
where $D_{B\mu}=\p_{\mu}-ie_BA_{B\mu}$ denotes the covariant derivative,
$F_{B\mu \nu }=\p_\mu A_{B\nu}-\p_\nu A_{B\mu}$ is the field strength,
and $\alpha$  a gauge parameter.
The bare character is indicated by the subscript `B.'
In principle, there are also ghost fields which, however, decouple
in the symmetric phase and remain massless.
Working in dimensional regularization they do not contribute to the energy
because of Veltman's rule,
\beq
\label{VR}
\int\f{d^Dp}{(2\pi)^D}\,p^\alpha =0\mbox{~~~for all~~~}\alpha.
\eeq

The coefficient $1/4$ in front of the coupling constant $g_B$ is
conventional.
The  Feynman diagrams associated with the vacuum energy of the Lagrangian
(\ref{EQqedlag}) have been generated iteratively in Ref.~\cite{recrel}.
At some places it will be useful to compare our results with those of an
earlier work \cite{twoloops}, where we have derived the two-loop effective
potential above and {\em below\/} $T_c$.
For such comparisons, a replacement $g_B\rightarrow 2g_B/3$ is required.

A full  extension of the work in Ref.~\cite{twoloops} is highly
nontrivial since the effective potential requires the calculation
of Feynman diagrams with three different masses.
For this reason we shall restrict ourselves in this paper to the symmetric
phase $T>T_c$, where the field expectations vanish and the system
contains only two masses, which greatly simplifies the problem,
in particular since one of the masses, the photon mass, is zero.
As a consequence, most diagrams can be reduced to scalar integrals which
can be computed exactly.
The only exception is the watermelon---or basketball---diagram whose
$\ep$ expansion is, however, known to sufficiently high order in $\ep$
\cite{chung3loops}.

As in Ref.~\cite{twoloops}, we shall use throughout Landau gauge
$\alpha\rightarrow 0$, which enforces a transverse photon field.
This has the advantage of being infrared stable \cite{Nogueira1,deCalan3}.

\section{Renormalization}
\label{calculation}
The renormalization constants of the model are defined by
\bea
\label{EQeren}
\phi_B=\phi\sqrt{\Zphi},~~~
A_{B\mu}=A_\mu\sqrt{\Za},~~~
m^2_B=m^2\f{\Zm}{\Zphi},~~~
g_B=g\mu^{\ep}\f{\Zg}{\Zphi^2},~~~
e_B=e\mu^{\ep/2}\f{\Ze}{\Zphi\sqrt{\Za}}=\f{e}{\sqrt{\Za}}\mu^{\ep/2},
\eea
where, in the last equation, we have taken into account the relation
$\Ze=\Zphi$, which is a consequence of the Ward identity.
Heuristically, this equality comes from the requirement that the covariant
derivative $D_{B\mu}\phi_B$ should not only be invariant with respect to
gauge transformations but also with respect to renormalization.
Thus it must acquire the same normalization factor as the field itself,
going over into $\sqrt{\Zphi}D_{\mu}\phi$.
An arbitrary mass scale $\mu$ in Eq.~(\ref{EQeren}) serves to define
dimensionless coupling constants $g$ and $e$.

The above multiplicative renormalizations are not sufficient
to extract all finite information from the theory.
The vacuum energy requires a special treatment, as emphasized in a previous
work of one of the authors (B.K.) \cite{ka96}.
Dimensionality requires the effective potential to have mass dimension $D$.
To have a finite vacuum energy we must add to the Lagrangian a counterterm
\beq
\label{LambdaB}
{E}_v^c=\f{m^4}{\mu^{\ep}}\Zv.
\eeq
The different renormalization constants may be expanded in powers of the
fluctuation size $\hbar$ as follows:
\beq
\label{EQrenconstexpansion}
Z_j=1+\sum_{l=1}^L \left[\f{\hbar}{(4\pi)^2}\right]^l
Z_j^{(l)},\qquad\qquad\Zv=\sum_{l=1}^L\left[\f{\hbar}{(4\pi)^2}\right]^l
\Zv^{(l)},
\eeq
where the subscript  $j$ stands for fields and coupling constants
$\phi,A,m,g,e$.
In minimal subtraction, each expansion coefficient has simple power series
\bea
\label{zj}
Z_j^{(l)}= \sum_{k=0}^l g^{l-k}e^{2k}\left( c_{j,k}^1\ep^{-l}+
c_{j,k}^2\ep^{1-l}+\dots +c_{j,k}^l\ep^{-1}\right),
\eea
except for $Z_g^{(l)}$, where the systematics is
\bea
\label{gzg}
gZ_g^{(l)}= \sum_{k=0}^{l+1} g^{l+1-k}e^{2k}\left(
c_{g,k}^1\ep^{-l}+ c_{g,k}^2\ep^{1-l}+\dots
+c_{g,k}^l\ep^{-1}\right),
\eea
and
\bea
\label{zv}
Z_v^{(l)}= \sum_{k=0}^{l-1} g^{l-1-k}e^{2k}\left(
c_{v,k}^1\ep^{-l}+ c_{v,k}^2\ep^{1-l}+\dots
+c_{v,k}^l\ep^{-1}\right).
\eea
Initially, one finds also pole terms of the form $1/\ep^2\times \ln$,
$1/\ep \times \ln$, and $1/\ep \times \ln^2$, where $\ln $ is short for
$\ln (m^2/\mubar^2)$ with $\mubar$ being related to the mass scale $\mu$
via the Euler-Mascheroni constant  $\ga_E$ as
$\mubar^2=4\pi\mu^2 \exp(-\ga_E)$.
These, however, turn out to cancel each other, which provides us with
a nice consistency check of the renormalization procedure \cite{SEEV}.

\section{Feynman rules and vacuum diagrams}
The elements of the Feynman diagrams associated with the Lagrangian
(\ref{EQqedlag}) are
\begin{fmffile}{graph1}
\bea
\parbox{22mm}{\begin{center}
\begin{fmfgraph*}(25,20)
\setval
\fmfleft{v1}
\fmfright{v2}
\fmf{fermion}{v1,v2}
\fmflabel{${\alpha}$}{v1}
\fmflabel{${\beta}$}{v2}
\end{fmfgraph*}
\end{center}}&=&\f{\delta_{\alpha\beta}}{p^2+m^2_B},\\
&&\nn\\
\parbox{22mm}{\begin{center}
\begin{fmfgraph*}(25,20)
\setval
\fmfleft{v1}
\fmfright{v2}
\fmf{photon}{v1,v2}
\fmflabel{${\mu}$}{v1}
\fmflabel{${\nu}$}{v2}
\end{fmfgraph*}
\end{center}}&=&\f{\delta_{\mu\nu}-p_{\mu}p_{\nu}/p^2}{p^2},\\
&&\nn\\
\parbox{22mm}{\begin{center}
\begin{fmfgraph*}(25,20)
\setval
\fmfleft{v1,v3}
\fmfright{v2,v4}
\fmfforce{0.5w,0.5h}{v5}
\fmf{fermion}{v1,v5}
\fmf{fermion}{v5,v2}
\fmf{fermion}{v5,v3}
\fmf{fermion}{v4,v5}
\fmflabel{${\alpha}$}{v3}
\fmflabel{${\delta}$}{v4}
\fmflabel{${\beta}$}{v1}
\fmflabel{${\ga}$}{v2}
\fmfdot{v5}
\end{fmfgraph*}
\end{center}}&=&-\f{g_B}{2}(\delta_{\alpha\beta}\delta_{\ga\delta}
+\delta_{\alpha\delta}\delta_{\beta\ga}),\\
&&\nn\\
\parbox{22mm}{\begin{center}
\begin{fmfgraph*}(25,20)
\setval
\fmfleft{v1,v3}
\fmfright{v2,v4}
\fmfforce{0.5w,0.5h}{v5}
\fmf{photon}{v4,v5}
\fmf{photon}{v2,v5}
\fmf{fermion}{v1,v5}
\fmf{fermion}{v5,v3}
\fmflabel{${\alpha}$}{v3}
\fmflabel{${\beta}$}{v1}
\fmflabel{${\mu}$}{v2}
\fmflabel{${\nu}$}{v4}
\fmfdot{v5}
\end{fmfgraph*}
\end{center}}&=&-2e^2_B\delta_{\alpha\beta}\delta_{\mu\nu},\\
&&\nn\\
&&\nn\\
\parbox{15mm}{\centerline{
\begin{fmfgraph*}(25,20)
\setval
\fmfleft{v3,v1}
\fmfright{v2}
\fmfforce{0.5w,0.5h}{v5}
\fmf{photon}{v2,v5}
\fmf{fermion}{v5,v1}
\fmf{fermion}{v3,v5}
\fmflabel{${\alpha,q_1}$}{v1}
\fmflabel{${\beta,q_2}$}{v3}
\fmflabel{${\mu}$}{v2}
\fmfdot{v5}
\end{fmfgraph*}
}}\quad&=&e_B\delta_{\alpha\beta}(q_1+q_2)_{\mu}.\\
\nn
\eea
\end{fmffile}

In Ref.~\cite{recrel}, the vacuum diagrams of the theory have been
generated recursively up to four loops.
Table~1 shows all diagrams needed for the three-loop vacuum energy
above $T_c$, omitting those which contain massless separable loop
integrals which vanish by Veltman's rule (\ref{VR}).

\section{Renormalization constants $\Zv^{(3)}$ and $\Zphi^{(2)}$}
\label{effpot}
The determination of the two-loop effective potential below $T_c$
\cite{twoloops} allows us to extract the following one- and two-loop
contributions to the renormalization constants.
Note that a factor $(4\pi)^{-2l}$ has been taken out in the definition
(\ref{EQrenconstexpansion}).
\bea
\label{EQZm1}
\Zm^{(1)}&=& g\f{(N+2)}{2\ep},
\\
\label{EQZg1}
g\Zg^{(1)}&=& \f{g^2(N+8)+48e^4}{2\ep},
\\
\label{EQZphi1}
\Zphi^{(1)}&=& e^2\f{6}{\ep},
\\
\label{EQZa1}
\Za^{(1)}&=&-e^2\f{N}{3\ep},
\\
\label{EQZv1}
\Zv^{(1)}&=& \f{N}{2\ep},
\\
\nn\\
\label{EQZm2}
\Zm^{(2)}&=&
\f{(N+2)}{\ep^2}\left[\f{1}{4}g^2(N+5)-3ge^2+6e^4\right]
-\f{1}{\ep}\left[\f{3}{8}g^2(N+2)-2ge^2(N+2)-e^4(5N+1)\right],
\\
g\Zg^{(2)}&=& \f{1}{\ep^2}\left[\f{1}{4}g^3(N+8)^2-
3g^2e^2(N+8)+12g e^4(N+8)+8 e^6(N+18) \right]
\nn\\
\label{EQZg2}
&&\mbox{} -\f{1}{\ep}\left[
\f{1}{4}g^3(5N+22)-2g^2e^2(N+5)-2ge^4(5N+13)+\f{4}{3}e^6(7N+90)\right],
\\
\label{EQZv2}
\Zv^{(2)}&=&
\f{N}{\ep^2}\left[-3e^2+\f{1}{4}g(N+2)\right]+2e^2\f{N}{\ep}.
\eea

\subsection{Results for the Feynman integrals}
In this section, we give the value of the diagrams listed in Table~1.
Although the exact value of part of the three-loop diagrams is known,
we only give the $\ep$ expansion through order $\ep^0$, for the sake of
brevity.
The notation is as follows: $I_{ln}$ is the integral for the case $N=2$,
omitting the weights of Table 1 and setting coupling constants and
scalar mass equal to unity.
The first index $l$ is the loop order while the second index $n$ counts
through the diagrams within each loop order as listed in Table 1.

We have with $D=4-\ep$
\bea
I_{1a}&=&-\f{1}{(4\pi)^{D/2}}\Gamma(-D/2),\\
I_{2a}&=&\f{1}{(4\pi)^D}
\left[\f{4\Gamma(2-D/2)}{3-D}-\Gamma(1-D/2)\right]\Gamma(1-D/2),\\
I_{2b}&=& \f{1}{(4\pi)^D} \Gamma(1 - D/2)^2 ,\\
I_{3a}&=&
\f{1}{(4\pi)^{6}}\left(\f{e^{\ga_E}}{4\pi}\right)^{-3\ep/2}
\Bigg[\f{4}{\ep^2}+\f{-\f{29}{3}+64\ze(3)}{\ep}
-\f{943}{12}+\f{64\ln^42}{3}+512{\rm Li}_4\left(\f{1}{2}\right)
\nn\\
\label{EqLi4}
&&\quad\quad\quad\quad\quad\quad\quad\quad\quad\mbox{}
+\f{3\ze(2)}{2}-128\ze(2)\ln^22+288\ze(3)-352\ze(4)\Bigg],\\
I_{3b}&=&\f{1}{(4\pi)^{6}}\left(\f{e^{\ga_E}}{4\pi}\right)^{-3\ep/2}
\left[\f{96}{\ep^3}+\f{242}{\ep^2}
+\f{\f{2701}{6}+36\ze(2)}{\ep}+\f{5945}{8}+\f{363\ze(2)}{4}+12\ze(3)
\right],\\
I_{3c}&=&\f{1}{(4\pi)^{6}}\left(\f{e^{\ga_E}}{4\pi}\right)^{-3\ep/2}
\left[-\f{208}{3\ep^3}-\f{188}{3\ep^2}-\f{179+26\ze(2)}{\ep}
-\f{2683}{12}-\f{47\ze(2)}{2}+\f{250\ze(3)}{3}\right],\\
I_{3d}&=&\f{1}{(4\pi)^{6}}\left(\f{e^{\ga_E}}{4\pi}\right)^{-3\ep/2}
\left[\f{12}{\ep^3}+\f{71}{2\ep^2}+\f{\f{1793}{12}
+9\ze(2)}{2\ep}+\f{12731}{96}+\f{213\zeta(2)}{16}
-\f{15\ze(3)}{2}\right],\\
I_{3e}&=&\f{1}{(4\pi)^{6}}\left(\f{e^{\ga_E}}{4\pi}
\right)^{-3\ep/2}
\left[-\f{24}{\ep^3}-\f{16}{\ep^2}-\f{34+9\ze(2)}{\ep}
-22-6\ze(2)+3\ze(3)\right],\\
I_{3f}&=&\f{1}{(4\pi)^{6}}\left(\f{e^{\ga_E}}{4\pi}\right)^{-3\ep/2}
\left[\f{8}{\ep^3}+\f{28}{3\ep^2}
+\f{5+3\ze(2)}{\ep}-\f{27}{4}+\f{7\ze(2)}{2}+7\ze(3)\right],\\
I_{3g}&=&\f{1}{(4\pi)^{6}}\left(\f{e^{\ga_E}}{4\pi}\right)^{-3\ep/2}
\left[\f{24}{\ep^3}+\f{40}{\ep^2}+\f{50+9\ze(2)}{\ep}
+56+15\ze(2)-3\ze(3)\right],\\
I_{3h}&=&0,\\
I_{3i}&=&\f{1}{(4\pi)^{6}}\left(\f{e^{\ga_E}}{4\pi}\right)^{-3\ep/2}
\left[\f{16}{\ep^3}+\f{92}{3\ep^2}+\f{35+6\ze(2)}{\ep}
+\f{275}{12}+\f{23\ze(2)}{2}-2\ze(3)\right],\\
I_{3j}&=&\f{1}{(4\pi)^{6}}\left(\f{e^{\ga_E}}{4\pi}\right)^{-3\ep/2}
\left[\f{8}{\ep^3}+\f{8}{\ep^2}+\f{6+3\ze(2)}{\ep}
+4+3\ze(2)-\ze(3)\right],
\eea
where ${\rm Li}_4$ in Eq.~(\ref{EqLi4}) denotes the polylogarithm of order 4:
${\rm Li}_4(x)=\sum_{k=1}^{\infty}x^k/k^4$ \cite{Grad}.

\subsection{Divergent terms through three loops}
Denoting by $E_{\rm vac}$ the vacuum energy density in the symmetric phase,
we find up to three loops the expansion
\beq
E_{\rm vac}=
\f{m^4}{\mu^{\ep}}\Zv +\hbar E_1 +\hbar^2 E_2 +\hbar^3 E_3
+{\cal O}(\hbar^4),
\eeq
with the one-loop part
\beq
E_1=m_B^D\f{N}{2}I_{1a},
\eeq
the two-loop part
\beq
E_2=m_B^{2D-4}\left[
\f{1}{2}e_B^2\f{N}{2}I_{2a}-\f{1}{2}g_B\f{N(N+2)}{8}I_{2b}\right],
\eeq
and the three-loop part
\bea
E_3
&=&
m_B^{3D-8}\bigg[
\f{1}{4}e_B^4 \f{N}{2}I_{3a}+\f{1}{2}e_B^4
\f{N}{2}I_{3b}+\f{1}{4}e_B^4\left(\f{N}{2}\right)^2I_{3c}-
e_B^4 N I_{3d}-\f{1}{2}e_B^4\f{N^2}{2}I_{3e}+\f{1}{4}e_B^4 2N I_{3f}
\nn\\
&&\quad\quad\quad\quad
-g_Be_B^2\f{N(N+2)}{8}I_{3g}+\f{1}{8}g_B^2\f{N(N+2)}{8}I_{3i}
+\f{1}{2}g_B^2\f{N(N+2)^2}{32}I_{3j}\bigg].
\eea

A few remarks are useful on the calculation of the pole terms of the
diagrams.
Taking into account the expansion of the renormalization constants in
the form (\ref{EQrenconstexpansion}), each Feynman integral is expanded
up to order $\hbar^3$.
The expansion coefficients are determined to cancel the $1/\ep^n$ terms
arising from the Feynman integrals.
In particular we have:

\begin{itemize}
\item[--]
To order $\hbar$, the cancellation of the $1/\ep$ pole leads directly
to the known  one-loop result (\ref{EQZv1}) as in Ref.~\cite{twoloops}.
\item[--]
To order $\hbar^2$, there are  pole terms of the form $1/\ep^2$, $1/\ep$
and $1/\ep\times$ $\ln$.
The $1/\ep\times$ $\ln$ terms cancel if
\beq
\label{EQZphione}
\Zphi^{(1)}= \Zm^{(1)}-\f{g(N+2)-12e^2}{2\ep},
\eeq
which is fulfilled by the one-loop expressions
(\ref{EQZm1}) and (\ref{EQZphi1}) for $\Zm^{(1)}$ and $\Zphi^{(1)}$,
respectively.
After this, the cancellation of the ordinary poles lets us recover
the result (\ref{EQZv2}) obtained before in Ref.~\cite{twoloops}.
\item[--]
Finally, we need to cancel the poles at the $\hbar^3$ level.
Besides poles without logs, there are poles of the types $1/\ep\times \ln$,
$1/\ep^2\times\ln$ and $1/\ep\times\ln^2$, which have to vanish.
To simplify the discussion, we introduce the notation $Z^{(i,j)}$ where
the first superscript $i$ indicates the loop order, and the second
superscript $j$ gives the order of the pole, i.e.,
$Z^{(i)}=\sum_{j=1}^i Z^{(i,j)}/\ep^j$.
\end{itemize}

When removing the pole proportional to $1/\ep\times  \ln^2$ we
obtain $\Zphi^{(2,2)}$ as a function  of $\Za^{(1,1)}$,
$\Zg^{(1,1)}$, $\Zm^{(1,1)}$ and $\Zm^{(2,2)}$.
Similarly, the removal of $1/\ep\times \ln$ gives $\Zphi^{(2,1)}$ as a
function of $\Za^{(1,1)}$ and $\Zm^{(2,1)}$.
Finally, the removal of the pole proportional to $1/\ep^2\times\ln$
gives $\Zg^{(1,1)}$ as a function of $\Za^{(1,1)}$ and $\Zm^{(1,1)}$.
Inserting this into the expression for $\Zphi^{(2,2)}$, we obtain
\beq
\Zphi^{(2,2)}=
e^4(6-5N)+6e^2\Zm^{(1,1)}
-\left[\f{3}{4}g^2(N+2)+\f{1}{2}g(N+2)\Zm^{(1,1)}
-\Zm^{(2,2)}\right].
\eeq
Taking into account all these relations, the $1/~\epsilon^n$ pole terms
are found to be removed with $\Zv^{(3,1)}$ and $\Zv^{(3,2)}$ which are
functions of $\Za^{(1,1)}$ only, while $\Zv^{(3,3)}$ is determined
independently of $\Za^{(1,1)}$.

Using the known results for $\Za^{(1,1)}$ and $\Zm^{(1,1)}$, we
recover the result (\ref{EQZg1}) derived before in Ref.~\cite{twoloops}.
Using the known result for $\Za^{(1,1)},\Zm^{(1,1)},\Zm^{(2,1)},\Zm^{(2,2)}$,
we also obtain
\bea
\Zphi^{(2,1)}&=&-\f{1}{16}g^2(N+2)-\f{1}{12}e^4(11N+18),\\
\Zphi^{(2,2)}&=&e^4(N+18),
\eea
this coinciding with previous results derived from a renormalization of
the two-point functions in Refs.~\cite{Tess,holo}.

Finally, inserting $\Za^{(1,1)}$ into $\Zv^{(3,1)}$ and $\Zv^{(3,2)}$,
we have
\bea
\Zv^{(3,1)}&=&
\f{N(N+2)g^2}{64}-\f{N[43N+294-384\ze(3)]e^4}{48},\\
\Zv^{(3,2)}&=&-\f{5N(N+2)g^2}{24}+2N(N+2)ge^2+\f{N(25N-38)e^4}{6},\\
\Zv^{(3,3)}&=&\f{N(N+2)(N+4)g^2}{8}-3(N+2)Nge^2+\f{N(5N+48)e^4}{3},
\eea
where, as mentioned above, $\Zv^{(3,3)}$ is independent of $\Za^{(1,1)}$.
For $e^2=0$ we recover the three-loop result of the pure $\phi^4$
theory of Ref.~\cite{ka96}.

\section{Renormalization group function of the vacuum $\gv$}
The critical behavior of the renormalization constant $Z_v$
is characterized by the finite renormalization group function
$\ga_v$ defined by the logarithmic derivative \cite{ka96}
\beq
\ga_v=-\f{\mu^{1+\ep}}{m^4}\f{dE_v^c}{d\mu}.
\eeq
Using standard methods \cite{VSF2}, $\ga _v$ can be extracted from the
simple pole terms of $Z_v$, whose residue will be denoted by $Z_v^{[1]}$,
as \cite{ka96}
\beq
\ga_v= \left(1+g\f{\p}{\p g}+\f{1}{2}e\f{\p}{\p e}\right)Z_v^{[1]}.
\eeq
The results of the last section for the $Z_v^{(l,1)}$ yield
\beq
\ga_v=\f{N}{2}\f{\hbar}{(4\pi)^2}
+4Ne^2\left[\f{\hbar}{(4\pi)^2}\right]^2
+\left\{\f{3N(N+2)}{64}g^2
+N\left[-\f{43}{16}N-\f{147}{8}+24\ze(3)\right]e^4\right\}
\left[\f{\hbar}{(4\pi)^2}\right]^3+{\cal O}(\hbar^4).
\eeq

\section{Vacuum energy density}
In the previous section, we have focused on the removal of
divergences, thus fixing the renormalization constants.
Since the three-loop integrals $I_{3a}$---$I_{3j}$ are known to zeroth
order in  $\ep^0$, we can also determine the finite vacuum energy
density of the symmetric phase of the Ginzburg-Landau model.
Up to a negative sign, it is given by the sum of the vacuum diagrams.
Having in mind the application of our result to phase transitions
in three dimensions, we give here its $\ep$ expansion.
We must calculate the one-loop diagram up to the order $\ep^2$ and the
two-loop diagrams up to  the order $\ep$.

The general form of the $L$-loop result has an $\ep$ expansion of the form
\beq
E_{\rm vac}=\f{Nm^4}{4\mu^\ep}
\sum_{l=1}^L\left[\f{\hbar}{(4\pi)^2}\right]^l
\sum_{k=0}^{L-l}\ep^kE_{lk},
\eeq
where we have assumed that $e^2$ and $g$ are of order $\ep$, which is
correct at the fixed point relevant for the neighborhood of the phase
transition.
The expansion coefficients are
\beq
E_{10}=\lna-\f{3}{2},
\eeq
\beq
E_{11}=-\f{1}{4}\lnb+\f{3}{4}\lna-\f{7}{8}-\f{\ze(2)}{4},
\eeq
\beq
E_{12}=\f{1}{24}\lnc-\f{3}{16}\lnb
+\left[\f{7}{16}+\f{\ze(2)}{8}\right]\lna
-\f{15}{32}-\f{3\ze(2)}{16}+\f{\ze(3)}{12},
\eeq
\beq
E_{20}=\left(\f{N+2}{4}g-3e^2\right)\lnb
-\left(\f{N+2}{2}g-14e^2\right)\lna +\f{N+2}{4}g-19e^2,
\eeq
\bea
E_{21} &=& \left(-\f{N+2}{8}g+\f{3}{2}e^2\right)\lnc
+\left[\f{3(N+2)}{8}g-\f{17}{2}e^2\right]\lnb
\nn\\
&&
+\left\{-\f{[4+\ze(2)](N+2)}{8}g+\f{44+3\ze(2)}{2}e^2\right\}\lna
+\f{[2+\ze(2)](N+2)}{8}g-\f{50+7\ze(2)}{2}e^2,
\nn\\
\eea
\bea
~~~~~~~~~~~ E_{30}
&=&\left[\f{(N+2)(N+4)}{16}g^2-\f{3(N+2)}{2}ge^2
+\f{5N+48}{6}e^4\right]\lnc
\nn\\
&& +\left[-\f{(N+2)(2N+15)}{16}g^2
+7(N+2)ge^2+\f{49N-438}{12}e^4\right]\lnb
\nn\\
&& +\left[\f{(N+2)(2N+39)}{32}g^2
-\f{23(N+2)}{2}ge^2+\f{-139N+290+384\ze(3)}{8}e^4\right]\lna
\nn\\
&&+\f{N+2}{192}g^2
+6(N+2)ge^2
\nn\\
&&+\left[
\f{1351N}{48}+\f{261}{8}
-\left(\f{56N}{3}+208\right)\ze(3)
+176\ze(4)
+64\ze(2)\ln^22-\f{32}{3}\ln^42
-256{\rm Li}_4\left(\f{1}{2}\right)
\right]e^4.
\eea

\section{Conclusion}
With the help of dimensional regularization and the modified
minimal subtraction scheme $\overline{\mbox{MS}}$ we have
computed the vacuum energy density in an $\ep$ expansion up to
three loops for the symmetric phase of the Ginzburg-Landau model.
Further, we have determined the renormalization group function
of the vacuum $\ga_v$, which is the same above and below $T_c$.
Both quantities will be needed for the calculation of the amplitude
ratios of the specific heat at the phase transition of the model.

To arrive at the final goal of deriving universal amplitude ratios for
the specific heat above and below the phase transition, we must perform
a similar calculation also in the ordered phase below $T_c$.
Such a calculation will be complicated by a proliferation of Feynman
diagrams, the appearance of more mass scales and infrared divergences
for $N>2$.
Fortunately, the amplitude ratio of the specific heat does not require
knowledge of  the full effective potential, but only its value at the
minimum, whose evaluation is simpler and will be given in future work.

Certainly it is hoped that a higher loop effective potential describing
the phase transition will give us specific information on the nature of
the superconductive phase transition, in particular on the value of the
Ginzburg parameter at which the transition becomes tricritical.

\begin{acknowledgments}
We thank Dr.~D.J.~Broadhurst, Dr.~J.-M.~Chung, and Dr.~B.K.~Chung for
helpful communications.
The work of B.VdB. was supported by the Alexander von Humboldt foundation
and the Institut Interuniversitaire des Sciences Nucl\'eaires de Belgique.
\end{acknowledgments}

\newpage

\begin{fmffile}{graph2}

\begin{table}[t]
\begin{center}
\begin{tabular}{|c|l|}
\hline\hline
\,\,\,$L,n_1,n_2,n_3$ &
\\
\hline
$1,0,0,0$ &
\rule[-10pt]{0pt}{25pt}
$1$
\parbox{22mm}{\begin{center}
\begin{fmfgraph*}(15,15)
\setval
\fmfleft{i1}
\fmfright{o1}
\fmf{fermion,right=1}{i1,o1}
\fmf{plain,right=1}{o1,i1}
\end{fmfgraph*}\end{center}}
\\
\hline
$2,0,0,2$ &
${\displaystyle \f{1}{2}}$
\parbox{22mm}{\begin{center}
\begin{fmfgraph}(20,20)
\setval
\fmfforce{0w,1/2h}{v1}
\fmfforce{1w,1/2h}{v2}
\fmf{fermion,right=1}{v2,v1}
\fmf{fermion,right=1}{v1,v2}
\fmf{photon}{v2,v1}
\fmfdot{v2,v1}
\end{fmfgraph}\end{center}}
\\
$2,1,0,0$ &
${\displaystyle \f{1}{2}}$\,
\parbox{22mm}{\begin{center}
\begin{fmfgraph}(35,35)
\setval
\fmfleft{i1}
\fmfright{o1}
\fmf{fermion,right=1}{i1,v1}
\fmf{fermion,right=1}{v1,i1}
\fmf{fermion,right=1}{o1,v1}
\fmf{fermion,right=1}{v1,o1}
\fmfdot{v1}
\end{fmfgraph}\end{center}}
\\ \hline
$3,0,0,4$ &
${\displaystyle \f{1}{4}}$
\parbox{22mm}{\begin{center}
\begin{fmfgraph}(17,17)
\setval
\fmfforce{0w,0h}{v1}
\fmfforce{1w,0h}{v2}
\fmfforce{1w,1h}{v3}
\fmfforce{0w,1h}{v4}
\fmf{fermion,right=0.4}{v1,v2}
\fmf{fermion,right=0.4}{v2,v3}
\fmf{fermion,right=0.4}{v3,v4}
\fmf{fermion,right=0.4}{v4,v1}
\fmf{boson}{v1,v3}
\fmf{boson}{v2,v4}
\fmfdot{v1,v2,v3,v4}
\end{fmfgraph}\end{center}}
${\displaystyle \f{1}{2}}$
\parbox{22mm}{\begin{center}
\begin{fmfgraph}(17,17)
\setval
\fmfforce{0w,0h}{v1}
\fmfforce{1w,0h}{v2}
\fmfforce{1w,1h}{v3}
\fmfforce{0w,1h}{v4}
\fmf{fermion,right=0.4}{v1,v2}
\fmf{fermion,right=0.4}{v2,v3}
\fmf{fermion,right=0.4}{v3,v4}
\fmf{fermion,right=0.4}{v4,v1}
\fmf{boson,right=0.4}{v1,v4}
\fmf{boson,left=0.4}{v2,v3}
\fmfdot{v1,v2,v3,v4}
\end{fmfgraph}\end{center}}
${\displaystyle \f{1}{4}}$
\parbox{22mm}{\begin{center}
\begin{fmfgraph}(17,17)
\setval
\fmfforce{-0.3w,1h}{v1}
\fmfforce{-0.3w,0h}{v2}
\fmfforce{1.3w,1h}{v3}
\fmfforce{1.3w,0h}{v4}
\fmf{fermion,right=1}{v2,v1}
\fmf{fermion,right=1}{v1,v2}
\fmf{fermion,right=1}{v4,v3}
\fmf{fermion,right=1}{v3,v4}
\fmf{boson}{v1,v3}
\fmf{boson}{v2,v4}
\fmfdot{v1,v2,v3,v4}
\end{fmfgraph}
\end{center}}
\\
$3,0,1,2$ &
%
$1$
\parbox{22mm}{\begin{center}
\begin{fmfgraph}(24,24)
\setval
\fmfforce{0w,0.5h}{v1}
\fmfforce{1w,0.5h}{v2}
\fmfforce{1/2w,0h}{v3}
\fmfforce{0.12w,0.8h}{v4}
\fmfforce{0.88w,0.8h}{v5}
\fmf{fermion,right=0.5}{v5,v4}
\fmf{fermion,right=0.6}{v4,v3}
\fmf{fermion,right=0.6}{v3,v5}
\fmf{boson}{v3,v4}
\fmf{boson}{v3,v5}
\fmfdot{v3,v4,v5}
\end{fmfgraph}\end{center}}
${\displaystyle \f{1}{2}}$
\parbox{22mm}{\begin{center}
\begin{fmfgraph}(20,20)
\setval
\fmfforce{-0.3w,1h}{v1}
\fmfforce{-0.3w,0h}{v2}
\fmfforce{0.8w,1/2h}{v3}
\fmfforce{1.3w,1h}{v4}
\fmfforce{1.3w,0h}{v5}
\fmf{fermion,right=1}{v1,v2}
\fmf{fermion,right=1}{v2,v1}
\fmf{plain,right=1}{v4,v5}
\fmf{fermion,right=1}{v5,v4}
\fmf{boson,left=0.2}{v1,v3}
\fmf{boson,right=0.2}{v2,v3}
\fmfdot{v1,v2,v3}
\end{fmfgraph}
\end{center}}
\\
$3,0,2,0$ &
${\displaystyle \f{1}{4}}$
\parbox{22mm}{\begin{center}
\begin{fmfgraph}(24,24)
\setval
\fmfforce{0w,0.5h}{v1}
\fmfforce{1w,0.5h}{v2}
\fmf{fermion,right=1}{v2,v1}
\fmf{fermion,right=1}{v1,v2}
\fmf{photon,left=0.4}{v1,v2,v1}
\fmfdot{v1,v2}
\end{fmfgraph}\end{center}}
\\
$3,1,0,2$ &
$1$
\parbox{22mm}{\begin{center}
\begin{fmfgraph}(20,20)
\setval
\fmfforce{0w,1h}{v1}
\fmfforce{0w,0h}{v2}
\fmfforce{1/2w,1/2h}{v3}
\fmfforce{4/4w,1h}{v4}
\fmfforce{4/4w,0h}{v5}
\fmf{boson}{v4,v5}
\fmf{fermion,right=1}{v1,v2}
\fmf{plain,right=1}{v2,v1}
\fmf{fermion,right=1}{v5,v4}
\fmf{fermion,right=0.4}{v4,v3}
\fmf{fermion,right=0.4}{v3,v5}
\fmfdot{v4,v5,v3}
\end{fmfgraph}
\end{center}}
${\displaystyle \f{1}{2}}$
\parbox{25mm}{\begin{center}
\begin{fmfgraph}(34,34)
\setval
\fmfforce{0w,2/3h}{v1}
\fmfforce{1/2w,2/3h}{v2}
\fmfforce{1w,2/3h}{v3}
\fmf{boson,right=1}{v1,v3}
\fmf{fermion,right=1}{v2,v1}
\fmf{fermion,right=1}{v1,v2}
\fmf{fermion,right=1}{v3,v2}
\fmf{fermion,right=1}{v2,v3}
\fmfdot{v2,v3,v1}
\end{fmfgraph}\end{center}}
\\
$3,2,0,0$ &
${\displaystyle \f{1}{8}}$
\parbox{22mm}{\begin{center}
\begin{fmfgraph}(22,22)
\setval
\fmfforce{0w,0.5h}{v1}
\fmfforce{1w,0.5h}{v2}
\fmf{fermion,right=1}{v2,v1}
\fmf{fermion,right=1}{v1,v2}
\fmf{fermion,left=0.4}{v1,v2}
\fmf{fermion,left=0.4}{v2,v1}
\fmfdot{v1,v2}
\end{fmfgraph}\end{center}}
${\displaystyle \f{1}{2}}$
\parbox{22mm}{\begin{center}
\begin{fmfgraph}(55,55)
\setval
\fmfforce{0w,1/2h}{v1a}
\fmfforce{1/3w,1/2h}{v1b}
\fmfforce{2/3w,1/2h}{v2}
\fmfforce{1w,1/2h}{v3}
\fmf{fermion,right=1}{v1a,v1b}
\fmf{fermion,right=1}{v1b,v1a}
\fmf{fermion,right=1}{v1b,v2}
\fmf{fermion,right=1}{v2,v1b}
\fmf{fermion,right=1}{v2,v3}
\fmf{fermion,right=1}{v3,v2}
\fmfdot{v1b,v2}
\end{fmfgraph}
\end{center}}
\\\hline\hline
\end{tabular}
\end{center}
\caption{Relevant one-particle irreducible vacuum diagrams
$W^{(L,n_1,n_2,n_3)}$ and their weights through three-loop order of
the $O(N)$ Ginzburg--Landau model, where $L$ denotes the loop order
and $n_1,n_2,n_3$ count the number of $g,e^2$ and $e$ vertices,
respectively.}
\end{table}

\end{fmffile}

\end{document}